\documentclass[a4paper]{article}

\usepackage{geometry}
\usepackage{graphicx, hyperref, setspace, amsmath, amssymb, titlesec, fancyhdr, multicol, parskip, indentfirst, etoolbox, caption, hyperref, xcolor}
\usepackage{cite}

\titleformat{\section}{\centering\large\scshape}{\thesection}{1em}{}
\titleformat{\subsection}{\normalsize\bfseries}{\thesubsection.}{1em}{}
\titlespacing{\section}{0pt}{6pt}{6pt}
\titlespacing{\subsection}{0pt}{6pt}{6pt}
\titlespacing{\subsubsection}{0pt}{6pt}{6pt}
\title{
    \textbf{Dressed to Gamble: How Poker Drives the Dynamics of Wearables and Visits on Decentraland's Social Virtual World} 
    \thanks{
        \sloppy
        \textbf{Cite (APA):} Trujillo, A., Bacciu, C., \& Abrate, M. (\the\year). Dressed to Gamble: How Poker Drives the Dynamics of Wearables and Visits on Decentraland's Social Virtual World. \textit{Journal of Metaverse, 5}(2), 145-155. https://doi.org/10.57019/jmv.1673676
    }
}

\date{} 
\renewcommand{\thesection}{\Roman{section}.}
\renewcommand{\thesubsection}{\textit{\Alph{subsection}.}}
\renewcommand{\thesubsubsection}{\textit{\arabic{subsubsection}.}}

\titleformat{\subsection}{\normalfont\large\itshape}{\thesubsection}{1em}{}
\titleformat{\subsubsection}{\normalfont\itshape}{\thesubsubsection}{1em}{}

\fancypagestyle{firstpage}{
    \fancyhead[C]{
        \centering
        {\fontsize{14pt}{10pt}\selectfont
        \textbf{Journal of Metaverse}\\
        \textbf{Research Article}}\\
        {\fontsize{8pt}{10pt}\selectfont
        \textbf{Received:} 2025-04-10 \textbf{Reviewing:} 2025-04-13 \& 2025-05-27 \textbf{Accepted:} 2025-06-02 \textbf{Online:} 2025-06-04 \textbf{Issue Date:} 2025-12-31}\\
        \textbf{Year:} 2025, \textbf{Volume:} 5, \textbf{Issue:} 2, \textbf{Pages:} 145-155, \textbf{Doi:} 10.57019/jmv.1673676
    }
}

\begin{document}

\maketitle
\vspace{-1.5cm}
\thispagestyle{firstpage}

\begin{multicols}{3}
    \centering
    \textbf{Amaury TRUJILLO}\\
    \textit{Istituto di Informatica e Telematica, Consiglio Nazionale delle Ricerche (IIT-CNR)}\\
    \textit{Pisa, Italy}\\
    \textit{amaury.trujillo@iit.cnr.it}\\
    \href{https://orcid.org/0000-0001-6227-0944}{%
    \includegraphics[width=10pt]{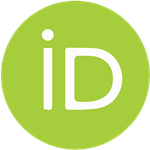}
    0000-0001-6227-0944
    }\\
    \textit{(Corresponding Author)}
    \vfill

    \textbf{Clara BACCIU}\\
    \textit{Istituto di Informatica e Telematica, Consiglio Nazionale delle Ricerche (IIT-CNR)}\\
    \textit{Pisa, Italy}\\
    \textit{clara.bacciu@iit.cnr.it}\\
    \href{https://orcid.org/0000-0002-8711-7812}{%
    \includegraphics[width=10pt]{orcid.png}
    0000-0002-8711-7812}    
    \vfill

    \textbf{Matteo ABRATE}\\
    \textit{Istituto di Informatica e Telematica, Consiglio Nazionale delle Ricerche (IIT-CNR)}\\
    \textit{Pisa, Italy}\\
    \textit{matteo.abrate@iit.cnr.it}\\
    \href{https://orcid.org/0009-0004-6809-3570}{%
    \includegraphics[width=10pt]{orcid.png}
    0009-0004-6809-3570}  
    \vfill
\end{multicols}

\singlespacing
\setlength{\parskip}{6pt}
\setlength{\parindent}{0.5cm}

\begin{multicols}{2}
\setlength{\columnsep}{0.5cm}

\noindent \textbf{\textit{Abstract---}Decentraland is a blockchain-based social virtual world where users can publish and sell wearables to customize avatars. In it, the third-party Decentral Games (DG) allows players of its flagship game ICE Poker to earn cryptocurrency only if they possess certain wearables. 
Herein, we present a comprehensive study on how DG and its game influence the dynamics of wearables and in-world visits in Decentraland.
To this end, we analyzed 5.9 million wearable transfers made on the Polygon blockchain (and related sales) over a two-year period, and 677 million log events of in-world user positions in an overlapping 10-month period.
We found that these activities are disproportionately related to DG, with its ICE Poker casinos (less than 0.1\% of the world map) representing a remarkable average share of daily unique visitors (33\%) and time spent in the virtual world (20\%). Despite several alternative initiatives within Decentraland, ICE Poker appears to drive user activity on the platform.
Our work thus contributes to the understanding of how play-to-earn games influence user behavior in social virtual worlds, and it is among the first to study the emerging phenomenon of virtual blockchain-based gambling.
}


\section{Introduction}
\label{sec:introduction}

In recent years, there has been much discussion around the Metaverse, an envisioned fully immersive iteration of the Internet, composed of a shared collection of interoperable 3D virtual worlds.
The Metaverse remains a vision in its early stages, but many social virtual worlds usually associated with it thrived during the \textsc{Covid-19} pandemic \cite{kerdvibulvech2022exploring}.
In these worlds, interaction with the environment and with other people is done primarily through an \emph{avatar}, an embodied representation of the user, with avatar customization being a significant in-world factor of purchase intent and one of the main pleasures for users~\cite{bleize2019factors}.
However, it has been argued~\cite{zhou2018ownership} that many of the ownership policies for in-world digital goods ---such as those used in avatar customization--- are disadvantageous to users. This is primarily due to the imbalance in rights between the users and platform proprietors. Traditionally, digital assets are not transferable independently of the platform; hence, these remain locked inside the virtual world or are forever lost when the underlying platform ceases to function.

In response, blockchain technology has been heralded by many tech enthusiasts as a decentralized and transparent approach to digital property in virtual worlds~\cite{cannavo2020blockchain}.
Arguably, Decentraland is the earliest and most popular blockchain-based virtual world, touted to be the first one ``built, governed, and owned by its users'' in its official website.
In Decentraland, users can transfer in-world land, usernames, and avatar wearables, either as part of a sale or for free. All these transfers are recorded on the Ethereum blockchain, or, in the case of wearables, on Polygon, a compatible and more scalable blockchain.

Based on these data, Trujillo and Bacciu~\cite{trujillo2023toward} were among the first to conduct a quantitative study of Decentraland's wearables.
They noticed that a single account ---related to the project \emph{Decentral Games} (DG)--- had by far the most published wearables.
DG manages a few in-world casinos, and possession of DG wearables is required to earn money (in the form of cryptocurrency) by playing \emph{ICE Poker}, its main attraction (depicted in Figure~\ref{fig:dg_casino_in_world}).
Owners can also \emph{delegate} their DG wearables to others, with whom earnings are split based on the wearable's level. 
Hence, DG wearables transcend the mere aesthetic in-world function of virtual garments, with ownership being enticing due to their earning potential.

Herein we thus delve into how DG and ICE Poker influence the dynamics of wearables' ownership and in-world visits in Decentraland. In particular, we seek to answer the following research questions (RQs):

\begin{description}
    \item[RQ1:] Are transfers of Decentraland wearables disproportionately related to those created by Decentral Games?
    \item[RQ2:] Do Decentraland users visit significantly more ICE Poker casinos compared to other in-world locations? 
\end{description}

To answer RQ1, we measure the share of wearables designed by DG in the overall dynamics of ownership transfers in different forms, including creation, sale, and destruction, as well as how the upgrading mechanism implemented by DG affects such transfers.
To answer RQ2, we analyze the visits ---both in time spent and number of visitors--- made on parcels in which ICE Poker casinos are located, compared to the rest of the in-world land.

\noindent
\begin{minipage}{\columnwidth}
    \centering
    \includegraphics[width=1\textwidth]{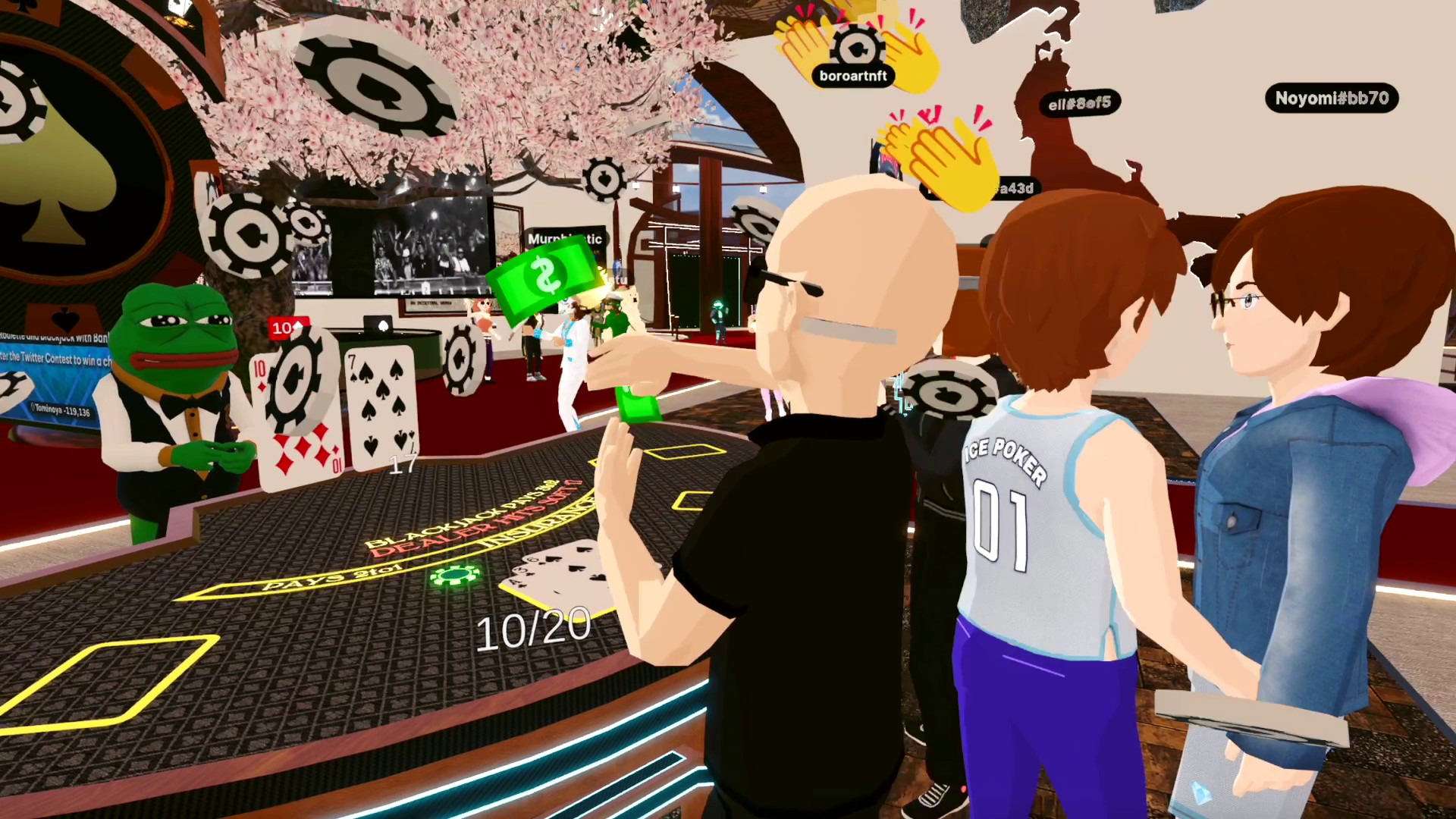}
    \captionof{figure}{Screenshot of an ICE Poker game session inside a Decentral Games casino.}
    \label{fig:dg_casino_in_world}
\end{minipage}

Our main contribution is therefore a quantitative analysis regarding the influence that online gambling has had on Decentraland's social virtual world.
In fact, integrating earning opportunities into virtual spaces has already resulted in serious repercussions in the real world\cite{grimmelmann2009virtual, axie}, emphasizing how virtual worlds can be both \emph{digital} (in their essence) and \emph{real} (in their effects) \cite{boellstorff2016whom}.
Thus, studying the use (and potential abuse) of Decentraland's affordances and economic incentives in the form of online gambling is important to better shape the future mechanisms and legal framework of social virtual worlds and the Metaverse in general.

\section{Background}
\label{sec:background}

In this section we briefly describe the main mechanisms, jargon, and events of both Decentraland and Decentral Games referred to in subsequent sections.

\subsection{Blockchain technologies}

Blockchains are decentralized, immutable ledgers that facilitate secure and transparent recording of transactions across a network of computers.
Decentraland is based on Ethereum~\cite{wood2014ethereum}, a versatile blockhain that enables \emph{smart contracts} ---self-executing programs that automatically enforce the terms of an agreement.
These provide the underlying infrastructure for non-fungible tokens (NFTs), unique identifiers often used to represent ownership of digital objects (e.g, art and collectibles).
Unlike fungible cryptocurrencies, NFTs are unique and non-interchangeable.
Smart contracts also enable decentralized autonomous organizations (DAOs), a form of token-based governance without a central authority, allowing for distributed decision-making and management~\cite{hassan2021decentralized}.

\subsection{Decentraland}

Publicly released in February 2020, Decentraland's social virtual world revolves around the so-called \emph{tokenomics}, i.e., an economy based on the exchange of assets via cryptographic tokens on blockchains~\cite{lo2020assets}.
Decentraland has its own cryptocurrency, \small\texttt{MANA}, with which users can buy in-world NFTs for digital assets such as land parcels and wearables.
Possession of \small\texttt{MANA} is also used as the basis of voting power for platform decisions via Decentraland's DAO~\cite{goldberg2023metaverse}.

The world (see Figure \ref{fig:decentraland_map_dg_casinos}) is composed of 90,601 square land parcels distributed on a 301$\times$301 grid.
Before the public release of the project, 37.4\% of the world map was allocated for districts, i.e., groups of adjacent parcels of privately owned land that share similar user interests.
Among the biggest districts we find, for example, the gambling-focused \emph{Vegas City} and the shopping-oriented \emph{Fashion Street}.

\noindent
\begin{minipage}{\columnwidth}
    \centering
    \includegraphics[width=1\textwidth]{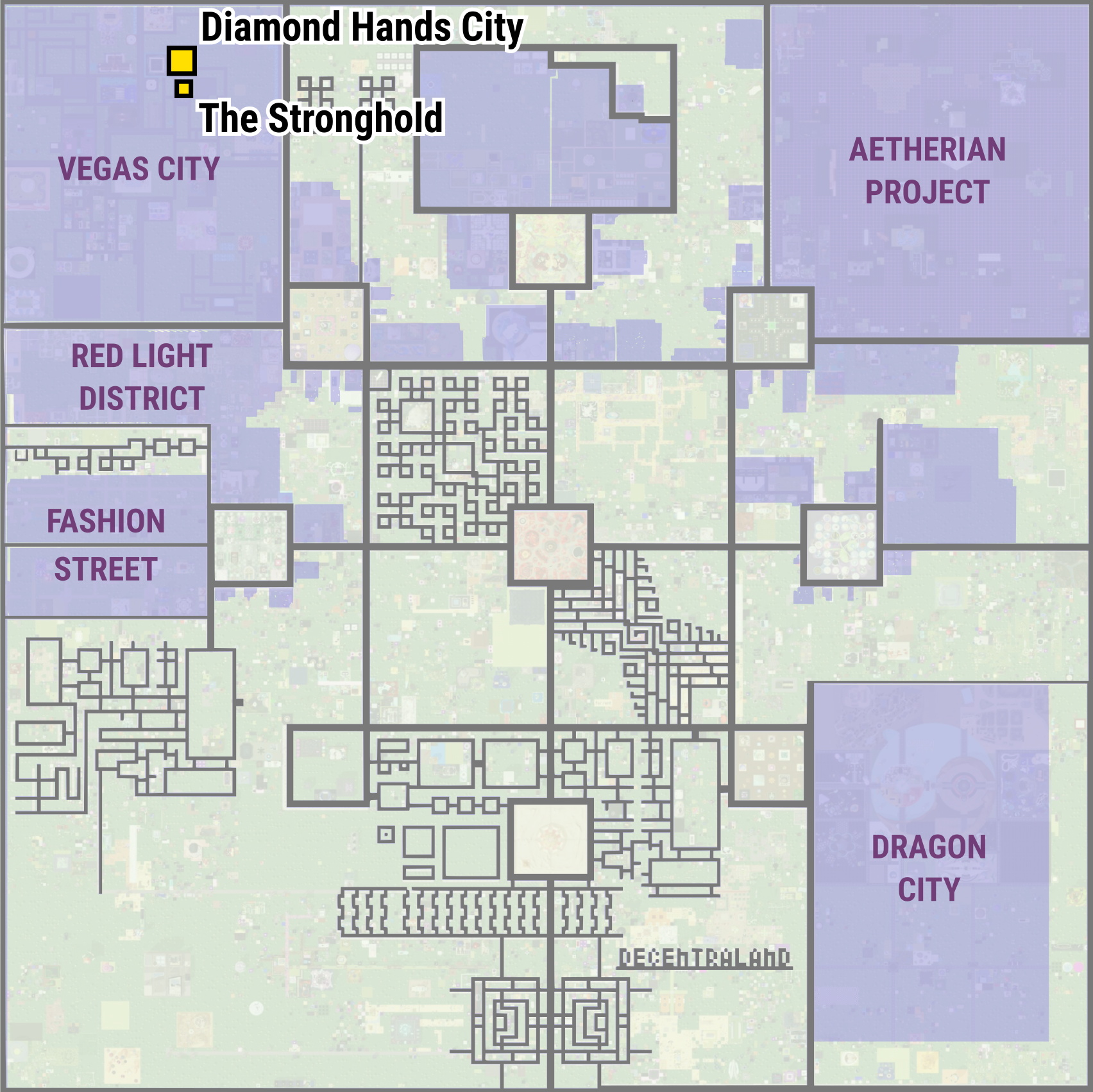}
    \captionof{figure}{Decentraland world map, where the biggest districts are indicated in violet and roads in gray. District \emph{Vegas City}  is home to \emph{Diamond Hands City} and \emph{The Stronghold} (both highlighted in yellow), ICE Poker casinos owned by Decentral Games. They cover only a minuscule surface of the map: less than 0.1\%.}
    \label{fig:decentraland_map_dg_casinos}
\end{minipage}

Decentraland wearables are virtual garments, accessories or full body costumes (called \emph{skins}) used to change the appearance of one's avatar. Wearables are grouped into collections of one or more items. Each item represents a distinct design from which one or more NFTs can be \emph{minted}, i.e, a new unique token ID representing the asset is created and assigned.
In addition, a given item has a rarity limit (from a predefined list) that sets its maximum supply of tokens ---e.g., a \emph{unique} item can be minted only once, while a \emph{common} item 100,000 times. To publish a wearable collection, creators pay a fee and submit their designs to the approval committee appointed by the DAO. Once approved, creators can sell their wearables via Decentraland's Marketplace (for a fee that goes to the DAO), or they can mint and transfer the respective NFTs to a blockchain address (their own or not), either as a gift or to be sold on a different marketplace.

\subsection{Decentral Games}

The project DG started in 2019 and has played a part in many initiatives in Decentraland ever since. In fact, DG established one of the first in-world casinos, with games such as roulette or poker, and it was a prominent creator of wearables since the beginning.
DG also established its own DAO, as well as its utility token and namesake \small\texttt{DG}, which grants access to rewards and voting power in the project's governance.
In October 2021, DG released ICE Poker, touted to be a Metaverse Play-to-Earn (P2E) game, i.e., one in which users are able to earn cryptocurrency simply by playing.
In preparation, DG also launched the cryptocurrency \small\texttt{ICE} to function as the in-game currency for rewards and minting wearables.
The following month, DG also published several exclusive ICE-themed wearables that granted attractive bonuses while playing the game, based on their ranking level.

Inside Decentraland, there are two DG virtual casinos dedicated to ICE Poker, both located in the Vegas City district. The first and more popular is called \emph{The Stronghold}; the second, \emph{Diamond Hands City}, is exclusive to ICE-themed wearables with the highest rank.
Outside Decentraland, DG wearables also give access to a related but separate mobile and web game released on September 2022, called \emph{ICE Poker Arcade}. In this version of the game, instead of cryptocurrency players earn electronic gadgets, DG merchandise or other virtual assets.

The level of access and/or reward provided by DG wearables varies by item, as illustrated by the examples in Figure~\ref{fig:dg_wearable_examples}; the more expensive the better. Within either game mode, players can \emph{wear} up to five wearables to augment their rewards. For certain ICE-themed wearables, it is also possible to upgrade their ranking level (ranging from 1 to 5) to gain better rewards by paying a smaller fee compared to buying one.
Additionally, all DG wearables can be delegated to other players, who can then split their earnings with the wearable's owner, with the latter retaining the rights over the asset on the blockchain and being able of revoking the delegation at any moment.

\noindent
\begin{minipage}{\columnwidth}
    \centering
    \includegraphics[width=1\textwidth]{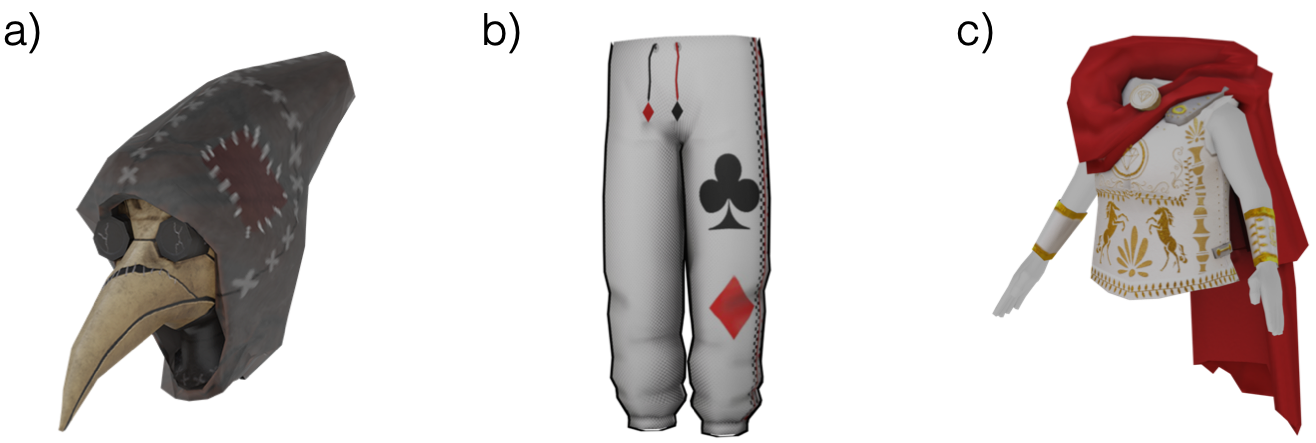}
    \captionof{figure}{Examples of wearables created by \textit{Decentral Games}: a) \emph{Venetian Mask (ICE Rank 1)}, which gives 1--7\% bonus; b) \emph{DG Suited Tracksuit Bottom}, which can be used to play Arcade Mode; and c) \emph{Paludamentum (ICE Rank 5)}, which gives 35--45\% bonus.}
    \label{fig:dg_wearable_examples}
\end{minipage}

\section{Related Work}
\label{sec:related_work}

While we are still in the early development stages of the Metaverse, numerous independent virtual worlds have already sprung up over the past few decades. Notable examples include Second Life, Meta Horizon Worlds, and massively multiplayer online (MMO) games such as World of Warcraft, Minecraft, Roblox and Fortnite.
However, these virtual environments have faced significant criticism regarding the governance and ownership of in-world assets. In Second Life, for instance, there has been a perceived emergence of virtual feudalism~\cite{grimmelmann2009virtual}.
Roblox has been criticized for profiting from child labor~\cite{parkin2022trouble} and for allowing harmful content and games with gambling-like mechanisms in its user-generated virtual worlds~\cite{kou2023harmful}. World of Warcraft has also grappled with conflicts between the game’s publisher and players over agency and ownership of in-game assets~\cite{glas2010games}.
More generally, there has been a separation between content and platform ownership in virtual worlds, often to the detriment of users, whose rights over bought or created in-platform content are usually restricted~\cite{zhou2018ownership}.

In response, a wave of blockchain-based platforms has emerged~\cite{cannavo2020blockchain}.
Virtual worlds such as Decentraland, The Sandbox, Axie Infinity, and Upland,
among others, attempt to mitigate the aforementioned issues by leveraging blockchain's transparency, security, and decentralization for digital ownership, cryptocurrency transactions, and governance. 
Decentraland~\cite{ordano2017blockchain} is a standout example, being the earliest and most popular virtual world based on blockchain. It has reached a wide audience and gone through many changes over several years, hence it offers a trove of publicly available data for the study of emerging phenomena on these novel asset-oriented platforms.
However, existing research on Decentraland has primarily focused on landholding assets~\cite{dowling2022fertile,goldberg2021land,guidi2022socialgames} or examined the platform’s tokens within the broader context of the NFT ecosystem~\cite{nadini2021mapping}.
There is a lack of studies regarding other kinds of digital assets available in Decentraland, namely avatar wearables,
with the latter having attracted the attention of the fashion industry~\cite{gonzalez2022digital}.

\subsection{Fashion in Virtual Worlds}
Fashion in social virtual worlds has been a subject of research for years, especially in Second Life, which has been active since 2003 and introduced the concept of customizable virtual spaces and avatars~\cite{bardzell2010virtual}. Despite its recent decline, Second Life maintained for years a robust economy centered on avatar customization, with many users becoming in-world fashion designers~\cite{boellstorff2015, boellstorff2020rethinking}.

Recently, the fashion industry’s attention has shifted towards popular MMO games, particularly through the creation of branded \textit{skins}, with an increasing interest in newer blockchain-based virtual worlds~\cite{rodriguez2022dressing}. Notably, several prestigious luxury brands are venturing into NFT collectibles~\cite{joy2022digital}, participating in events such as the Metaverse Fashion Week or designing wearables for Decentraland~\cite{gonzalez2022digital}, as well as retailing and engaging in other kinds of customer engagement within its social virtual world~\cite{goldberg2023retailing}.
In fact, Decentraland’s wearables are mainly distributed for free to boost engagement in other cryptoasset or Metaverse initiatives such as virtual casinos, with only a minor portion of them (3.4\%) being sold on the platform’s marketplace~\cite{trujillo2023toward}.

\subsection{Poker and Online Gambling}
The chance-based nature of the game of poker appeals to a wide audience of casual players, while the element of skill, particularly evident over many hands, entices professional players~\cite{fiedler2009quantifying}.
The introduction of the Internet to gambling activities has provided easy access for the general public, round-the-clock access, shorter intervals between bets, and instant reinforcements~\cite{griffiths2008internet}.
After 2003, online poker experienced tremendous growth, with a total market value reaching 3 to 4 billion USD by 2010, with approximately 5.5 million players participating in virtual poker rooms across the globe~\cite{fiedler2011market}.
Among such players, a subset has emerged as professionals, especially adept at \emph{multitabling} ---simultaneously playing multiple games to further reduce the element of chance. As reliance on poker earnings grew, these players developed a strong extrinsic motivation to play, even if not entirely removed from the gameplay. Further, it has been found that playing for real money strongly influences both game dynamics and player experiences in poker games~\cite{zaman2014motivation}.

Notably, professional gamblers share psychological traits with problem gamblers, highlighting the complex interplay between skill and risk~\cite{newall2023elite}.
Problem gamblers, often drawn to low-stakes gambling, may develop distorted risk perceptions and ultimately underestimate financial risks in general~\cite{armstrong2018exploration}.
Interestingly, players are socially motivated to gamble by the virtual communities they form, while these same communities also serve as safeguards against excessive behavior~\cite{sirola2021role,savolainen2022online}.
Online poker’s unique subculture ---which includes websites, forums, and chat rooms from all around the world--- fosters cooperative behavior among players through the sharing of information and strategies. Nonetheless, the game’s inherent competitiveness persists and is reflected in the hierarchical structures within these global communities~\cite{oleary2013online}.

\subsection{Crypto Asset Trading}
Cryptocurrency has emerged as a disruptive force in the financial landscape since Bitcoin's introduction in 2008.
Nonetheless, its trading remains characterized by high volatility, often leading to substantial crashes followed by significant rebounds~\cite{fang2022cryptocurrency,oecd2022lessons}.
NFTs entered the scene in 2014, gaining widespread attention in early 2021 due to very high-profile sales and causing considerable hype~\cite{houser2022navigating}.
NFT buyers exhibit a profound enthusiasm for the foundational technologies that facilitate unique avenues for creative expression and the development of innovative content creation business models. However, these positive sentiments are counterbalanced by the contentiousness of digital ownership, prevalence of low-quality NFTs, risk of scams, potential money laundering activities, and evolving legal framework~\cite{sharma2022s}.

There is also an overlap of personality and demographic traits among problem gamblers and cryptocurrency traders~\cite{johnson2023cryptocurrency}. In both cryptocurrency trading and gambling, decisions are often based on limited information, short-term motives for gain, and with high uncertainty; hence, people attracted to the latter are likely to be attracted to the former~\cite{delfabbro2021cryptocurrency}. Numerous operators in the crypto gambling sector provide access to a variety of betting options, games, and online casinos. However, these platforms often pose a significant risk as they are easily accessible to minors and vulnerable individuals, and typically offer minimal to no consumer protection. Further, they still remain largely unregulated in many countries~\cite{andrade2023safer}.

\subsection{Play-to-Earn Games}
Blockchain technology has also boosted a form of online gaming known as ``Play-to-Earn'' (P2E), where players receive monetary rewards for their gameplay, mainly in the form of the game's own cryptocurrency. Axie Infinity ---a Pokémon-style battle game with collectible axolotl-like creatures and in-game tokenomics--- is arguably one of the most well-known examples.
Its popularity peaked in 2021, particularly in the Philippines and later Venezuela, where it represented a viable source of earnings for low-income people during the height of the \textsc{Covid-19} pandemic~\cite{de2022play}.
However, it is known that extrinsically motivated gameplay, i.e., mainly driven by external rewards such as financial gain, may reduce intrinsic motivation and consequently have negative implications for player experience~\cite{delic2024profiling,hong2023wean}.
In fact, a survey conducted with a group of Axie Infinity players in 2020--21 has shown that, while some players enjoyed social interactions, overall intrinsic motivation tended to be low, causing high amounts of stress and unwanted tension to reach daily quotas~\cite{de2022play}.
This assessment was later corroborated by a study that analyzed online conversations about the game~\cite{delic2024profiling}.

Interestingly, part of the initial success of Axie Infinity was due to delegation models that emerged organically within the community, called \emph{scholarships}~\cite{de2022play}. These allow players to share their knowledge or delegate their gameplay (and assets) to others for a share of the rewards.
Nonetheless, while P2E games offer players from less-advantaged countries earning prospects above the national average wage, these might find themselves in exploitative schemes to provide low-cost labor for wealthier owners of NFTs~\cite{delic2024profiling}.
In lieu of the community-based scholarships of Axie Infinity, ICE Poker integrates elements of P2E games by design, allowing players to earn tokens just by participating in certain events, and providing them with sanctioned asset delegation.
Wealthy owners are thus offered an official mechanism to delegate their DG wearables to willing players for whom the initial investment is perhaps prohibitively expensive.

\section{Methods}
\label{sec:methods_and_data}

For our study, we focus on the two years after the launch of Decentraland wearables on Polygon, that is, from June 2021 to May 2023.
For this, we undertake a data-driven approach based on public data available from blockchains and related datasets, focusing on obtaining macro-level patterns of actual user behavior related to our RQs.
In the following, we describe the overall approach to answer our RQs and how we performed data collection.

\subsection{Overall approach}

For RQ1 (wearable transfers) we first retrieved all wearables created during the period of study, marking those created by DG based on metadata from Decentraland. We then retrieved from the Polygon blockchain the respective transfers, as well as sales and wearable upgrades from related sources. We then analyzed the proportion between DG wearables and the rest in terms of the number of transfers and sales volume (both in number and USD).

For RQ2 (in-world visits to ICE poker casinos), we first collected in-world tracking data of Decentraland users, available only from 12 July 2022. Hence, we analyzed the last 10 months of the two-year period of interest, in terms of the number of accounts and time spent visiting the parcels in which ICE poker casinos are located, compared to the rest of Decentraland's map.

Our descriptive and statistical analyses mostly concern proportions, i.e., the share of wearables or land associated with DG compared with the rest of Decentraland.
To test for significant differences in proportions we used two-sided tests with significance level $\alpha = 0.05$.
When analyzing samples we used $\chi^2$ tests, either goodness-of-fit test or association test according to the case; when analyzing populations we used a binomial test; for both kinds of test we use relative risk (RR) to measure effect sizes.

\subsection{Data collection}

For wearable metadata (e.g., collection, category, creator), we used several of the openly available  Decentraland's query endpoints on TheGraph, a service for querying blockchain networks, and whose definitions are open source.\footnote{\url{https://github.com/decentraland}}
For transfers, we collected the respective transactions from Polygon via the blockchain web service Alchemy.
For sales, we collected data from two different sources: 1) Decentraland's Marketplace for internal sales, using the aforementioned TheGraph query endpoints; and 2) OpenSea, the largest overall NFT marketplace and main means for external sales, as well as the initial default marketplace for DG wearables.

\noindent
\begin{minipage}{\columnwidth}
    \centering
    \includegraphics[width=1\textwidth]{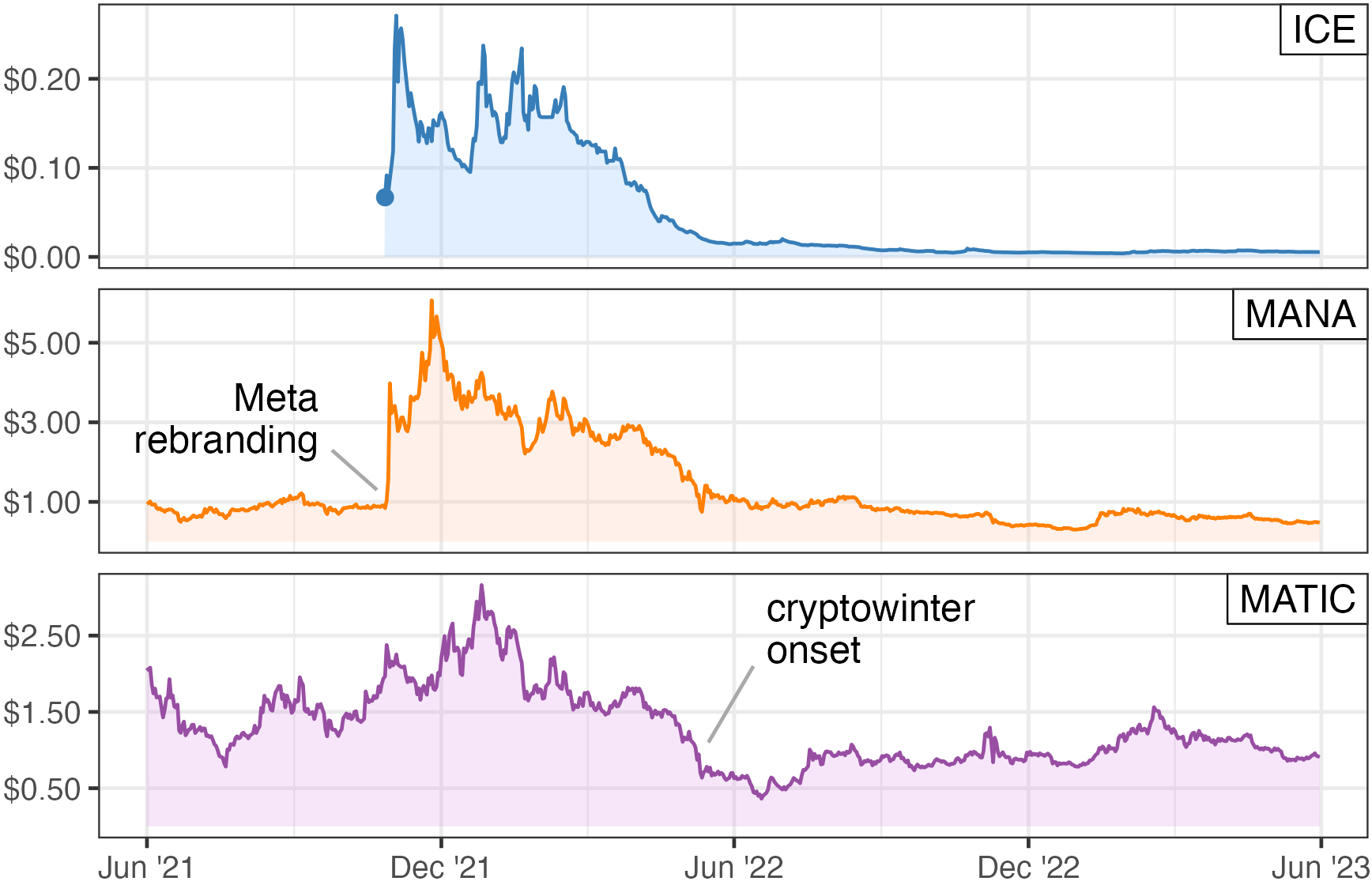}
    \captionof{figure}{Cryptocurrency price in USD during the two-year period of study for: \small\texttt{ICE} by Decentral Games, \small\texttt{MANA} by Decentraland, and \small\texttt{MATIC} by Polygon.}
    \label{fig:token_prices}
\end{minipage}

We also retrieved from Yahoo! Finance the price of the few cryptocurrencies used in the collected wearable sales, which we adjusted for inflation.
Price changes of these cryptocurrencies also serve to gauge interest on the studied platforms.
As can be seen in Figure~\ref{fig:token_prices}, the \small\texttt{ICE} token was released around the rebranding of Facebook into Meta in November 2021, with the latter provoking a sudden price increase in Decentraland's \small\texttt{MANA} (fourfold in just a couple of days). This was a considerable surge, even taking into account the already upward trend of \small\texttt{MATIC}, Polygon's cryptocurrency. However, the collapse of several interconnected cryptocurrencies the following May marked the onset of the 2022 cryptowinter~\cite{oecd2022lessons}, which affected a large share of blockchain projects and lasted until mid 2023.

For data on wearable upgrades, we again used Alchemy to retrieve the log events of the DG's smart contract on Polygon\footnote{Address \texttt{0xC9a67eD1472A76d064C826B54c144Ca00DAE4015}} that managed such transactions.
We then used the aforementioned wearable metadata to derive the ranking level and rarity limit of the upgraded tokens.

For in-world user tracking, we utilized the corresponding data archives from Atlas Data Corp, a blockchain analytics company that obtained a grant~\footnote{\url{https://decentraland.org/governance/project/?id=7422a99e-8a25-4625-9b30-ab688de5dade}} from Decentraland's DAO to keep record of user sessions. These data are log events (created every circa 20 seconds) of the current location of every connected user on the several servers that manage the 3D rendering of in-world scenes and avatars on Decentraland.
Given that this system started operating in mid July 2022, we narrowed our analysis of in-world visits to only 10 months, from August 2022 to May 2023. Moreover, we take into consideration only users with an Ethereum-compatible address, that is, we discard the small minority of guests not associated with an account.
We then transformed the more than 677 million log events for this time frame into sessions with a start and end for each land parcel and account address. Finally, we segmented the respective sessions' duration on a daily basis, adding 10 seconds to each to compensate for possible shortenings due to the 20-second polling rate of the log events.

\section{Analyses and Results}
\label{sec:results}

\subsection{RQ1: Overall wearable transfers}
\label{sec:transfers_and_sales}

From June 2021 to May 2023, a total of 4,629 collections with 8,450 wearable items (i.e., designs) were published by 1,517 distinct creators. Of these, we identified and manually verified two accounts\footnote{Addresses \texttt{0x17a253c2ac0d5ba92cadbbf665e3390c9913dc5d} and \texttt{0x00e5d44f6a296c10f159486f842838bd68f13e32}} that created all of DG wearables, based on marketplace offerings, item names, and user profiles on Decentraland.
Both DG accounts (0.1\% of creators) published 117 collections with 723 items (8.6\% of the total available items). Henceforth, we refer to these as DG wearables.

During these 24 months, there were 5,904,191 wearable transfers in total, with the vast majority (78.2\%) being mints (i.e., newly created NFTs recorded on the blockchain). For all transfers, the share of DG wearables was 15.1\%, while for mints it was 15.7\%.
Based on a binomial test, the proportion of minted DG wearables is significantly higher ($p\ll.01$) with a relative risk (RR) of 1.04 (i.e., 4\% higher than expected), compared to their share in total transfers. In other words, DG wearables are significantly less likely to be transferred once minted compared to other wearables, but the effect size is relatively small.
At most circa 59 million wearables could be minted based on the items' rarity limit, of which 5.3\% correspond to DG wearables, hence this 15.7\% mints share was significant ($p\ll.01$) and remarkable (RR = 2.95) when considering the maximum supply limit of tokens.

Only 3.4\% of the transfers made were part of a sale on Decentraland's Marketplace or OpenSea. For these, DG wearables represented merely 0.9\% of the total count. Based on the aforementioned transfers share of DG wearables (15.1\%) and a $\chi^2$ goodness-of fit-test, the share difference for number of sales is significant ($p\ll.01$), with a RR of 0.061 (93.9\% lower than expected based on the number of transfers).
Nevertheless, in terms of sales volume in dollars, DG wearable sales (again, 0.9\% in number) represented 17.3\% of the total 4,895,498\,USD. In other words, despite being considerably less in number of sales, the volume of money involved in the sales of DG wearables was remarkably higher compared to the rest of wearables, with a RR of 19.2 (circa 18 times higher than expected based on the number of sales).

In our analysis on a monthly basis, we noticed great variation in the share of DG wearables over time, but in different ways for transfers and sales. On one hand, and as illustrated in Figure~\ref{fig:monthly_dg_transfers}, the total volume of transfers reached its highest values in March and April of 2022, which coincides with the first edition of the Metaverse Fashion Week at the end of March. In those two months the share of DG wearables was only 3.3\% and 4.5\%, respectively. However, the transfers monthly share of DG wearables increased noticeably from September 2022 onward, reaching its peak in May 2023 with 85.3\%, while the rest of wearable transfers decreased in number. This DG share increase is mostly related to a few wearable items with a \emph{common} rarity limit and which give access only to ICE Poker Arcade, whose app was released precisely at the end of September.
On the other hand, for the monthly sales volume in USD (illustrated in Figure~\ref{fig:monthly_dg_sales_usd}), the highest monthly value for both total volume and share of DG wearables (62.9\%) is November 2021, the month in which the first DG collections focused on ICE Poker were released for minting. Moreover, the timing also coincides with the wake of Meta's rebranding, which provoked a surge and sustained high price in both \small\texttt{ICE} and \small\texttt{MANA} for that month, as shown in Figure~\ref{fig:token_prices}.

\noindent
\begin{minipage}{\columnwidth}
    \centering
    \includegraphics[width=1\textwidth]{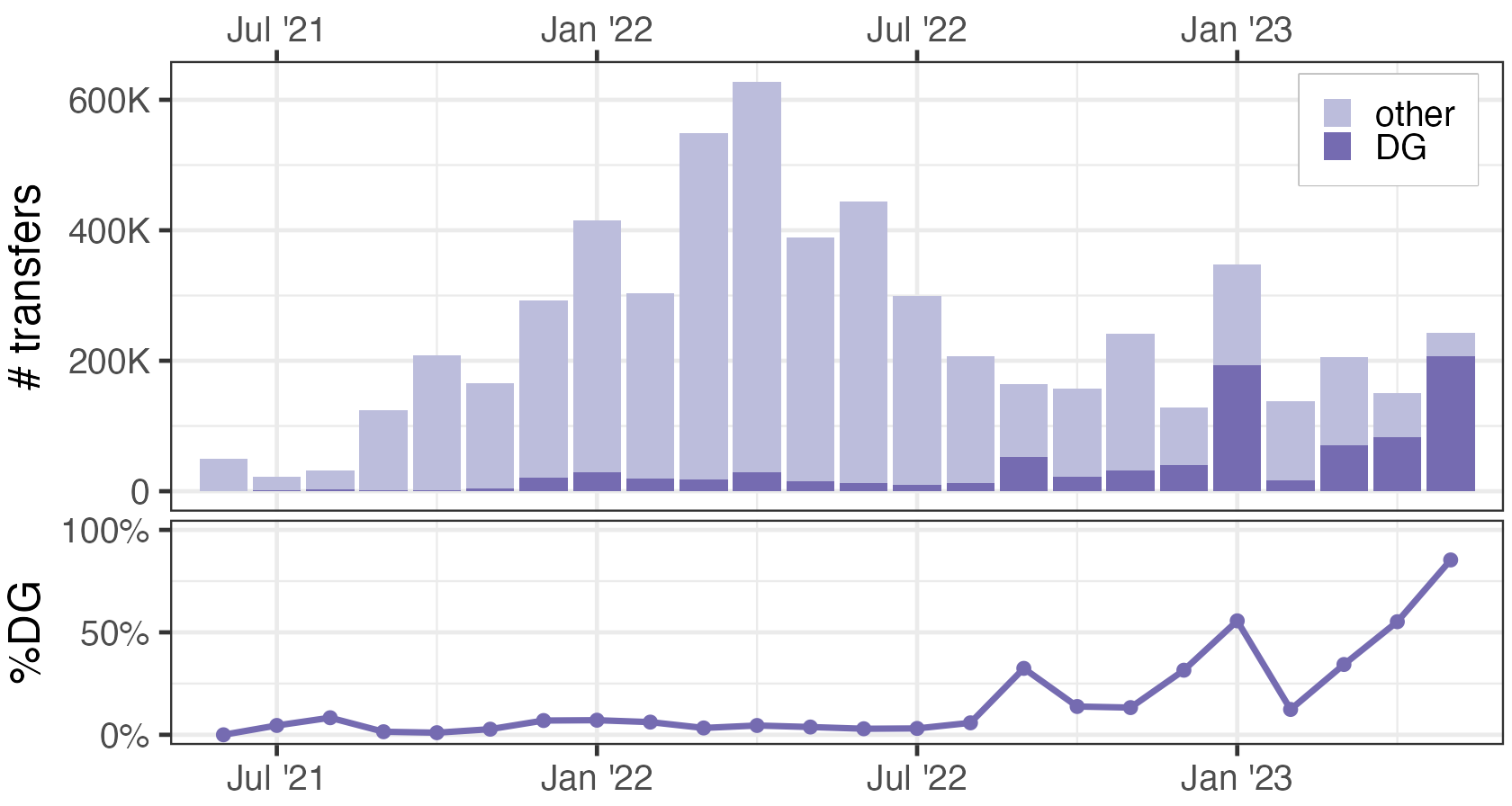}
    \captionof{figure}{Monthly number of transfers of Decentraland wearables on the Polygon blockchain and the respective Decentral Games (DG) share.}
    \label{fig:monthly_dg_transfers}
\end{minipage}

For the two-year period there were 58,791 mints of wearables with an ICE ranking level, of which  42,390 (72.1\%) were via upgrades.
Each upgrade implies three wearable NFT transfers on Polygon: 1) transfer of the original NFT from the owner to the upgrading smart contract; 2) \emph{burning} of this NFT by transferring it to a special \emph{dead} address, effectively removing it from circulation; and 3) \emph{minting} a new NFT for the owner from a higher-ranking wearable item within the same collection.
Hence, there were 127,170 transfers related to upgrades, representing 14.2\% of DG wearable transfers and 2.1\% of all transfers.
Interestingly, only 10\% of the mints of ICE-ranking wearables with a level higher than 1 was not related to an upgrade. In other words, 90\% of owners preferred to burn a lower-ranking token and mint a higher-ranking token via an upgrade, rather than buy (or wait to win) one without sacrificing the former.

\noindent
\begin{minipage}{\columnwidth}
    \centering
    \includegraphics[width=1\textwidth]{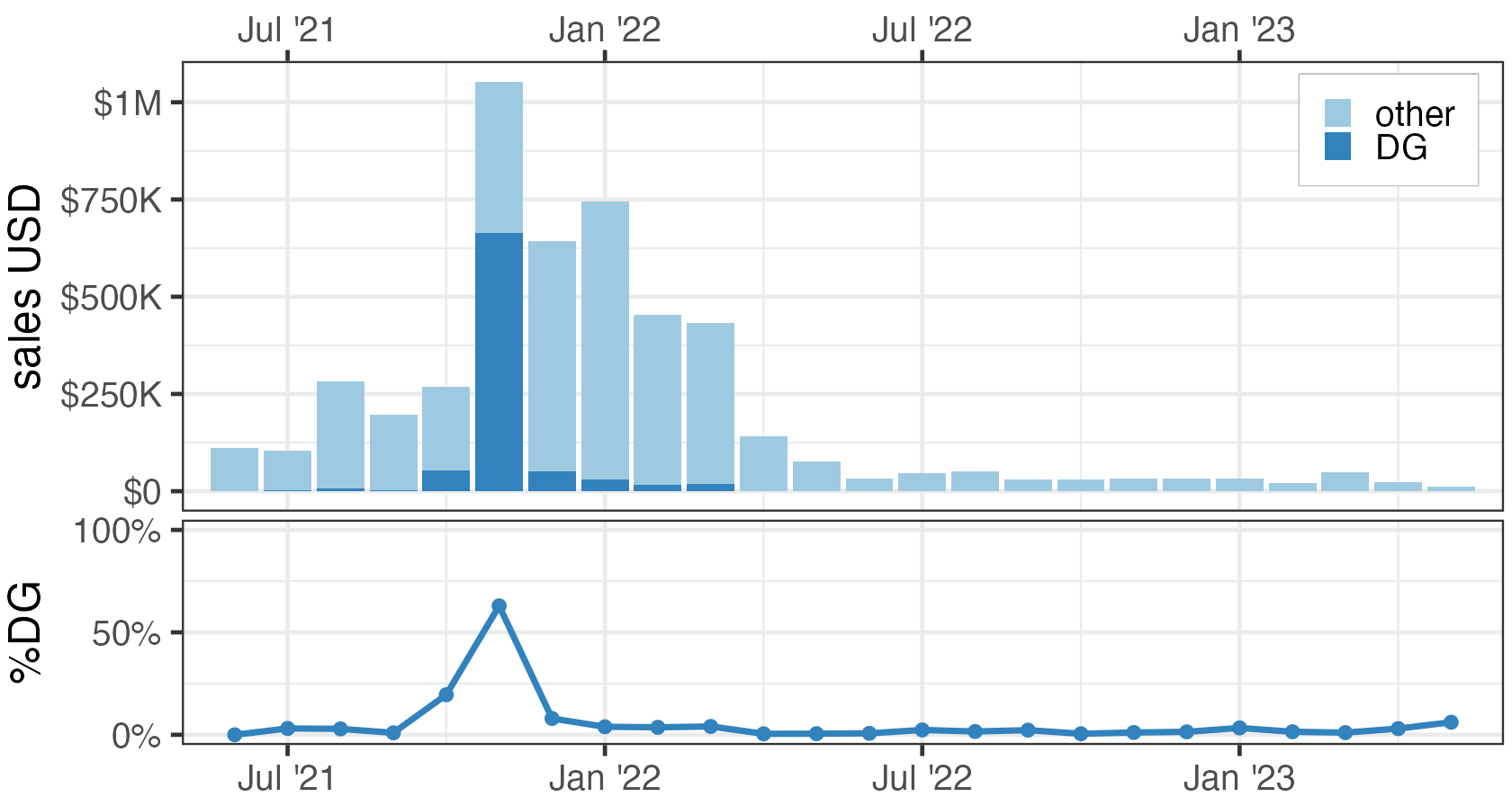}
    \captionof{figure}{Monthly sales volume in USD of Decentraland wearables ---made via the marketplaces of Decentraland and OpenSea--- and the respective share associated with Decentral Games (DG).}
    \label{fig:monthly_dg_sales_usd}
\end{minipage}

\subsection{RQ2: In-world visits of poker casinos}

Of the 90,601 land parcels in the world map, the two ICE Poker casinos cover 65 parcels, which is less than 0.1\% (see Figure~\ref{fig:decentraland_map_dg_casinos}), with \emph{The Stronghold} having only 16, and the exclusive \emph{Diamonds Hand City} 49.
For the ten months of log events of user in-world location we analyzed, there were a total of 451,413 distinct account addresses whose avatars collectively spent circa 7.5 million hours on Decentraland's world map. Of this time, 22.7\% was spent on the two ICE Poker casinos. By far, most time was spent on \emph{The Stronghold}, with 22.5\% of the total time, having a striking RR of 1278.5 ($p\ll.01$) based on land share; whereas \emph{Diamond Hands City} only represented 0.2\% of the total time, having a considerable RR of 3.5 ($p\ll.01$).
Furthermore, the parcels of \emph{The Stronghold} at the casino's official coordinates (-100, 127) and the adjacent (-101, 127) were by far the most visited single parcels in the whole world map, respectively with 6.6.\% and 5.8\% of the total time spent on Decentraland.

When analyzed on a daily basis, visits to the ICE Poker casinos are subject to relatively small fluctuations ---compared to the rest of the world map--- both in terms of time spent 
(see Figure~\ref{fig:daily_visits_hours}) and number of distinct visitor accounts (see Figure~\ref{fig:daily_visits_accounts}).
For visits to ICE Poker casinos, we have a moderate but steady downward trend for both time spent and visitors.
In the rest of the world map, the trend is slightly downward for time spent, while the trend is visibly downward for visitors.
In other words, over time there were less distinct accounts visiting Decentraland, but these collectively spent more time on it.
However, upon further inspection this increase in time spent per visitor is mostly due to a few accounts with a much higher number of hours than average.
Indeed, in terms of share, time spent on ICE Poker casinos reached its highest levels at the beginning of September 2022, up to 53\%, but it decreased to circa 20\% in May 2023 due to the aforementioned outlier users in the rest of the world map. Nevertheless, for visitors, the share at the beginning of the ten-month period was of circa 32\%, and at the end it increased to circa 41\%.

\noindent
\begin{minipage}{\columnwidth}
    \centering
    \includegraphics[width=1\textwidth]{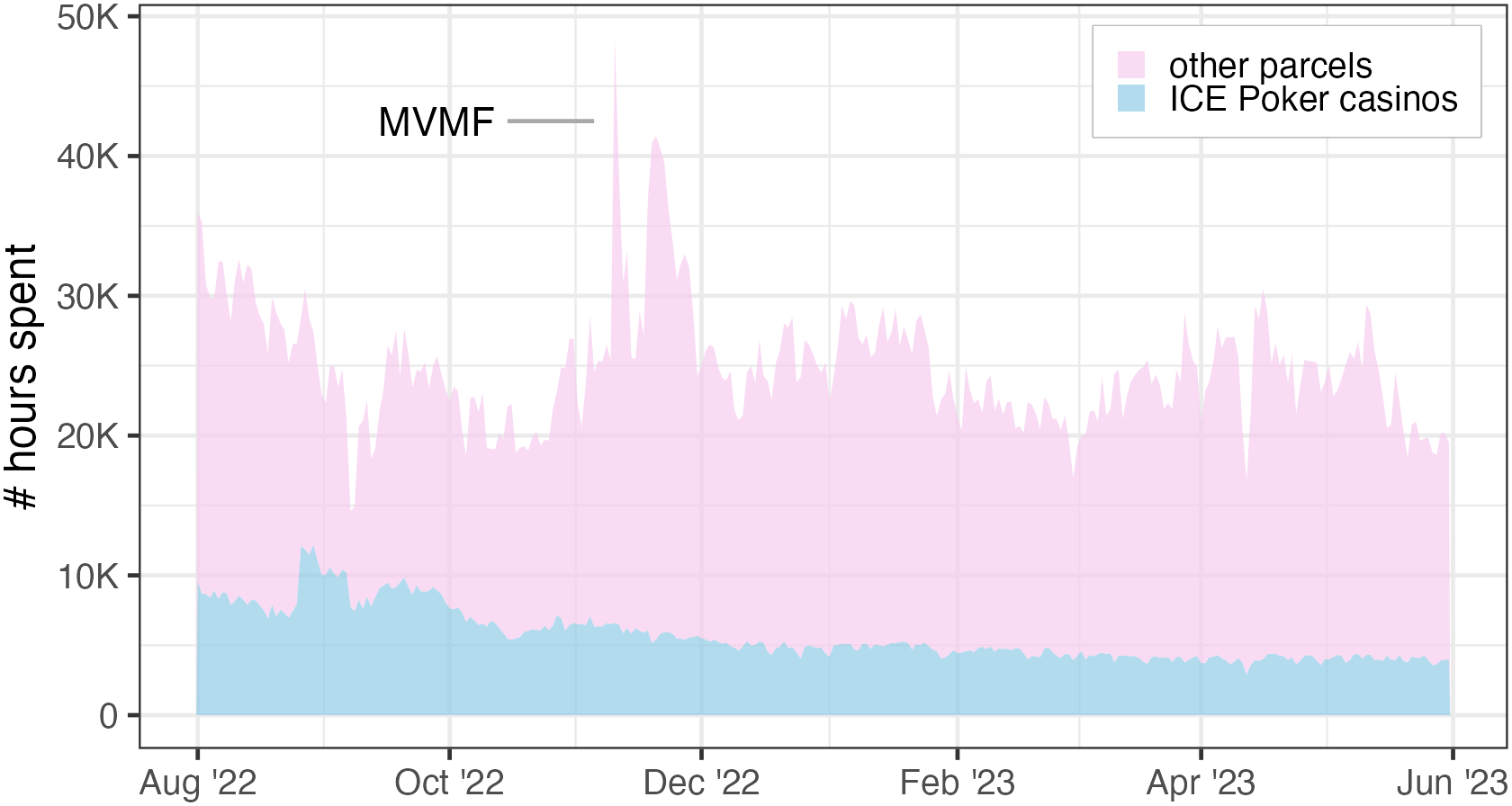}
    \captionof{figure}{Daily time spent (in hours) visiting Decentraland. ICE Poker casinos had a remarkable share, with a daily median of 20\%. The highest overall spike was reached during the Metaverse Music Festival (MVMF) in mid-November. }
    \label{fig:daily_visits_hours}
\end{minipage}

\noindent
\begin{minipage}{\columnwidth}
    \centering
    \includegraphics[width=1\textwidth]{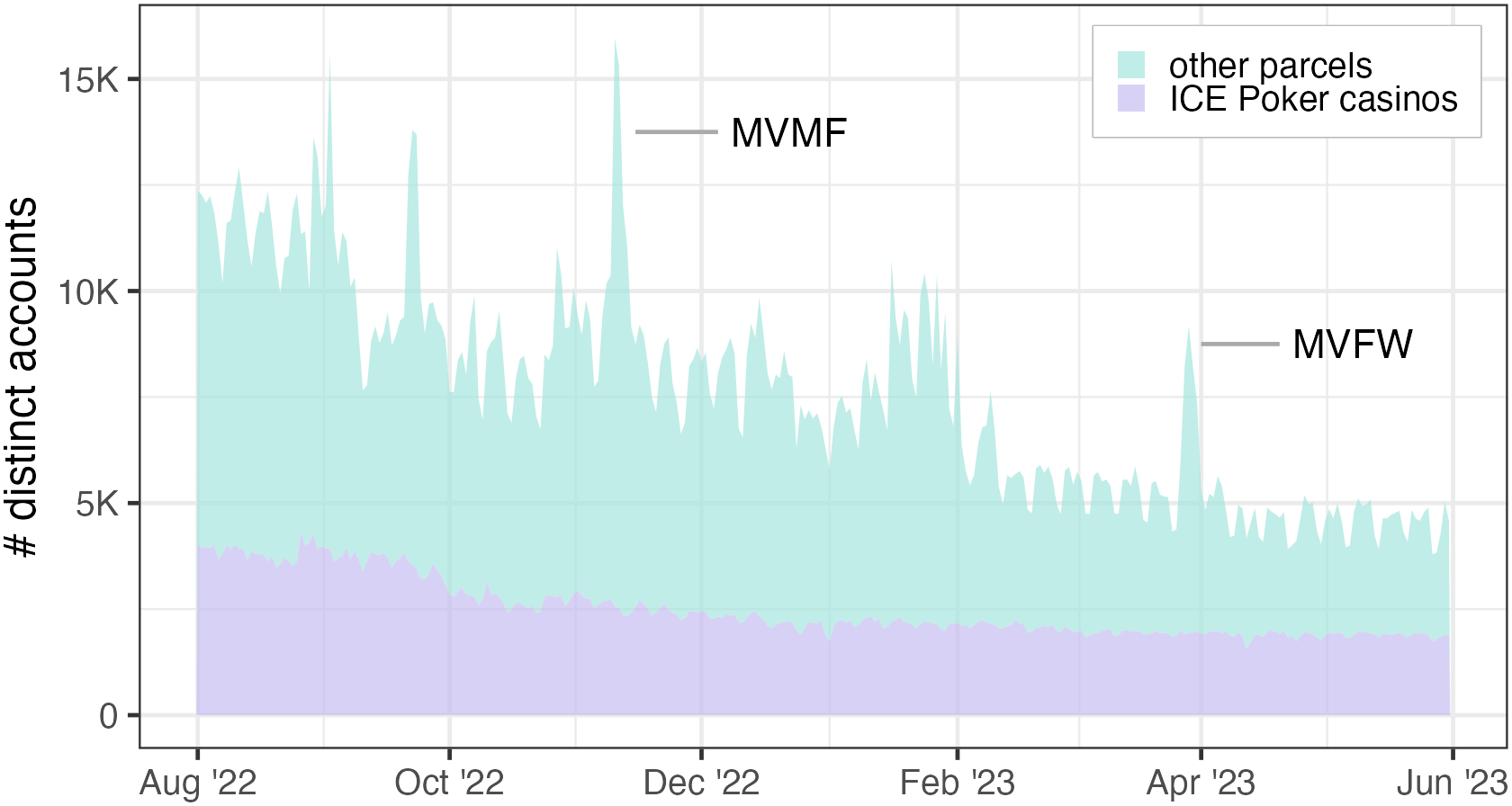}
    \captionof{figure}{Daily distinct accounts that visited Decentraland. ICE Poker casinos had a striking share, with a daily median of 33\%. The Metaverse Music Festival (MVMF) in mid-November had the highest spike overall, while the Metaverse Fashion Week (MVFW) at the end of March had the only spike during the last four months.}
    \label{fig:daily_visits_accounts}
\end{minipage}

On one hand, visits to the ICE Poker casinos had no spikes nor dips. There was, however, a short-lived surge (for both time and accounts) that started at the end of August 2022, when DG gifted special wearables to circa 3,000 of the most active players as a reward and in preparation for its new Tournament game mode.
On the other hand, in the rest of the world map there were several visit spikes, surges, and dips. In particular, the Decentraland Metaverse Music Festival, held on 10--13 November 2022, was the event that provoked the most visits (both in time and accounts).
Incidentally, the Metaverse Fashion Week of 2023 also provoked a spike in visitors at the end of March, but not in time spent and only during a period of four months in which the number of distinct visitors had already remarkably declined.

\section{Discussion}
\label{sec:discussion}

Based on our results, we can respond affirmatively to both RQ1 and RQ2: wearable transfers on Decentraland are indeed disproportionately related to those created by DG, and ICE Poker casinos do receive significantly (and remarkably) more visits than other in-world locations.
In the following, we discuss our main findings, we reflect upon the ambiguous legal standing of ICE Poker, we explore the social implications of our findings, and finally, we describe the main limitations of our study.

\subsection{Main findings}

Our findings confirm that DG and ICE Poker have great influence on Decentraland wearable and visit dynamics.
Just two DG accounts out of 1,517 wearable creators published 8.6\% of the total available items, with these being twice more likely to be minted based on the maximum supply limit compared to the rest, particularly at the end of the period of study. DG wearables also had a monetary sales volume an order of magnitude higher despite the considerably fewer sales. Moreover, in spite of covering less that 0.1\% of the land, ICE Poker casinos had a striking share of time spent and unique visitors in Decentraland, with a respective daily median of 20\% and 33\%. Further, the vast majority of these visits were done to a single casino: \emph{The Stronghold}.
Still, these dynamics have varied over time, on account of the initiatives that both Decentraland and DG have carried out to attract and retain users, and due to external market forces.

The Metaverse Fashion Week is arguably the most important initiative around Decentraland wearables. Both editions of 2022 and 2023 attracted well-established fashion brands, were featured in mainstream media, and piqued the interest of the general public. The initiative, however, did not meet expectations.
Based on a qualitative study on the 2023 edition~\cite{miikkulainen2023digital}, most user complaints were related to on-boarding complexity, a lackluster in-world user experience, the context switch between in-world and traditional web-based interaction, and the need of paid approval even for wearables designed for oneself.
In our analysis of visits we indeed detected a spike of distinct visitors during the 2023 edition at the end of March (see Figure~\ref{fig:daily_visits_accounts}), but there was no visible effect on the time spent on the platform (see Figure~\ref{fig:daily_visits_hours}), and the number of visitors continued its downward trend afterward.
The fact that in 2024 there was no Metaverse Fashion Week confirms the failure of the initiative to keep up the momentum going.
On the other hand, the Metaverse Music Festival attracted much more visitors that spent considerable time in Decentraland (see Figures~\ref{fig:daily_visits_hours} and~\ref{fig:daily_visits_accounts}), but nevertheless visits did not visibly increase in the following weeks.
It remains to be seen if this initiative will continue to gather interest.

As for DG, the main initiative after the release of ICE Poker wearables was the introduction of Tournament mode, which changed the rules to win prizes and claim \small\texttt{ICE} tokens. These changes were deemed necessary due to the quick pace with which delegates, who mostly did not own other kinds of DG assets (e.g., wearables or its utility token), withdrew \small\texttt{ICE} to convert it to other cryptocurrency or fiat money, accelerating its inflation.
In order to not alienate core delegates with these changes, DG gifted wearables to the 3,000 most active ones, which seems to explain the surge in visitors and time spent in ICE Poker casinos at the end of August 2022 (see Figures \ref{fig:daily_visits_hours} and~\ref{fig:daily_visits_accounts}).
Another important DG initiative was the launch of its Arcade mode and related web and mobile apps in September 2022, which allow players to earn non-monetary prizes without the need of visiting Decentraland; possession of DG wearables is still necessary, however. This may explain the increase in the share of transfers of DG wearables at the end of the period of study (see Figure~\ref{fig:monthly_dg_transfers}) without the corresponding increase in visits share to ICE Poker casinos.

Concerning external market forces, the \small\texttt{ICE} token and the initial ICE Poker wearables were by happenstance released a few days shy of Meta's rebranding.
This image change of such a behemoth corporation is considered to be the main factor behind the surge in popularity of the term Metaverse and the hype thereof~\cite{dolata2023metaverse}.
Consequently, the valuation of DG's newly introduced cryptocurrency and wearables reached prominent heights, along with Decentraland's own cryptocurrency, as can be seen in Figures~\ref{fig:token_prices} and~\ref{fig:monthly_dg_sales_usd}.
Nevertheless, the initial hype regarding NFTs has since subsided (also that of the Metaverse), with attention having shifted to generative artificial intelligence, which came to the spotlight with the release of ChatGPT~\cite{echauri2025tech}.

\subsection{On the legality of ICE Poker}

After a few years of uncertainty regarding its legal status, online gambling is now banned or heavily regulated in most jurisdictions~\cite{zborowska2012regulation}. Similarly, for many years and due to their novelty, cryptoassets remained in a regulatory limbo in most jurisdictions, but after many instances of scams and fraud ---as in the case of the exchange FTX--- in the last few years there has been a push to outright ban or closely regulate cryptocurrencies~\cite{trautman2022ftx}.
ICE Poker is a combination of both gambling and cryptocurrencies. Hence, this begs the simple question on whether this kind of operation is legal or not. The answer is, however, murky.

Firstly, on the project's front page it was stated that DG's parent company operated in Costa Rica. This is most likely due to the country's \textit{laissez-faire} approach for both online gambling and cryptocurrencies ~\cite{kerr2018can}, and the fact that its operating licenses are often used for online gambling services in foreign jurisdictions in which licensing is more rigid or non-existing~\cite{fischer2021poker}.
Moreover, in the official project documentation it is stated that the company developing DG, Web4 LTD, is established in the British Virgin Islands, a British overseas territory known for being a tax haven and having links to dubious online gambling operations~\cite{macegoniuk2021british}.

Curiously, the terms and conditions of ICE Poker state that is not covered by gambling laws in many jurisdictions due to the lack of (fiat) money exchange therein.
Based on the number of chips and wearables they possess, players periodically receive \small\texttt{ICE}, which can ultimately be converted to fiat money, but on an external exchange such as Binance.
Yet, players must guarantee they are not residents of one of the dozens of non-supported regions listed on the project's website due to their strict gambling and cryptocurrency regulations.

\subsection{Social implications}

Platform transparency and openness are stated to be paramount in Decentraland. The project strives toward these qualities by means of its use of public records on the blockchain, user-oriented ownership of in-world assets, public and open-source code repositories, governance via a DAO, and initiatives to overtly monitor the ecosystem ---e.g., by financing the collection of in-world tracking data we used to analyze visits. These characteristics, common to many blockchain-based projects, also offer unprecedented opportunities for both industry and academia to develop and understand novel sociotechnical systems in the form of virtual reality spaces, but not only~\cite{cannavo2020blockchain}. Nevertheless, the current levels of transparency and openness in Decentraland do not preclude the abuse of the platform.

For instance, in June 2022 DG submitted a proposal to Decentraland's DAO for a grant,\footnote{\url{https://decentraland.org/governance/proposal/?id=99c66b90-e2d2-11ec-9000-175d8dd584b8}} asking the equivalent to one million USD in \small\texttt{MANA} to increase its liquidity pool for \small\texttt{ICE} rewards. The grant was approved by 56\% of the voting power, which is based on ownership of Decentraland \small\texttt{MANA}, usernames, and land holdings. DG ---one of the earliest and most prominent actors in the platform--- accounted for almost a third of the voting power in its own favor.
This is an illustrative example of how a decentralized governance is not necessarily a neutral one. Rather, in its current form Decentraland's DAO gives way to automated situations reminiscent of the \emph{principal–agent problem}~\cite{goldberg2023metaverse}, because ``what is best for a dominant individual is not necessarily a good choice for the virtual world as a whole as well as the majority of its stakeholders''.

Other examples of potential platform abuse are the aforementioned drastic reward changes introduced with Tournament mode and the revamping of ICE Poker itself. These highlight that despite the assurances of public transactions and smart contracts, most of the functionality of the poker system ---and the virtual world itself--- lives outside the blockchain, and is thus subject to arbitrary changes by its controlling entity.

In addition, a third-party actor was able to exploit the affordances provided by the virtual world, which lead to deviations in user behavior unforeseen by the platform's creators. The decision by DG to use Decentraland wearables as access tokens to play in its online casino games has significantly influenced the overall acquisition trends of wearables at the platform level, perhaps tilting the motivating factors for acquiring wearables from intrinsic (e.g, the aesthetic enjoyment of the garment) to extrinsic (e.g., the earning utility of the token).
Hence, functional wearables might be appreciated solely for their utility, whereas merely cosmetic ones might be deemed less valuable due to their absence of utility, which weakens Decentraland's claim to be a social space primarily focused on creativity and self-expression.
In fact, an overabundance of extrinsic motivation can often undermine an experience by eroding the spontaneous intrinsic motivation, especially for games~\cite{hong2023wean}.

Based on our results, in-world visits are also greatly related to DG casinos. As with wearable transfers, this phenomenon is likely influenced by the earning potential of ICE Poker, in addition to the ludic enjoyment alone that casinos and other world areas offer.
Moreover, ICE Poker combines gambling and cryptoasset trading, with potentially troubling consequences to players due to self-destructive behavior, such as engaging in risk-taking and thrill-seeking activities, and a susceptibility to addiction~\cite{johnson2023cryptocurrency}.

Lastly, professional gamblers in ICE Poker and ICE Arcade could potentially replicate exploitative practices observed in other Play-to-Earn (P2E) games. They might use scholarships and delegation as a form of digital feudalism, where the majority of the profits are funneled to a small group of individuals or entities. This could lead to increased social disparity and potentially exacerbate economic inequalities~\cite{zaucha2024racial}, particularly if low-income individuals are involved, similar to the situation observed in Axie Infinity~\cite{de2022play}.

The study of the emerging phenomena in novel platforms such as Decentraland and games such as Ice Poker is increasingly important. A better understanding of their underlying mechanisms and dynamics will allow stakeholders to be better informed and prepared to enact the necessary changes to tackle these and other related issues that might emerge in the next few years.
In particular, we invite policymakers to take a critical look into the growing phenomenon of gambling and gambling-like activities in social virtual worlds based on tokenomics. Given the risk-taking behavior associated with both cryptocurrencies and gambling, together with their novel use in immersive virtual worlds, consumer protection should be a priority, especially for vulnerable individuals, such as those who are both low-income and younger. To this end, policymakers could follow an approach similar to the recent tackling of abuse of loot boxes in video games~\cite{xiao2024loot}, by expanding the legal definition of gambling and adapting existing regulation, such as enforcing the disclosure of probabilities regarding risks and rewards, and applying age rating and warning labels.

In view of the above legal and economic uncertainties, we also suggest developers of social virtual worlds to be wary of heavily relying ---directly or indirectly--- on such gambling-like activities to keep users engaged. Moreover, particular care should be taken in designing and implementing the decentralized governance and affordances of the platform, as to not allow a single third party to gain an overtly dominant role. Indeed, Decentraland's reliance on ICE Poker represents a predicament for the platform, which already suffers from a volatile ecosystem and diminished interest, raising doubts on its long-term stability. In this regard, at the end of 2024 the project announced a new version of the platform, Decentraland 2.0, with the stated goals of strengthening social interactions, boosting user engagement, improving discoverability, and increased immersion and fun. At the moment of writing, it remains unclear if this new version will succeed in reaching its goals and if it will overcome the dominance of in-world gambling activities.

\subsection{Limitations}

In spite of our careful approach, several limiting factors could influence the interpretation and applicability of our results. In particular, despite having public access to all of the wearable transactions on Polygon, we had to make use of sample data for information not readily available in order to better comprehend both transfers and visits.

One such limitation concerns sales data, as we only collected data from two marketplaces. Tracking every single transfer that is also a sale is burdensome because each marketplace has its own smart contracts and sales protocol. Moreover, these might even change over time.
Nonetheless, the marketplaces we selected should cover the vast majority of wearable sales, as Decentraland's Marketplace is the official and default venue, and OpenSea is the largest NFT marketplace and was the initial default venue to sell wearables by DG, which now has its own marketplace.

Another potential limitation regards in-world tracking of users. We did not use the term \emph{active users} in our analysis because we only considered the presence of a given account address in the log events, i.e., we did not filter out potential bots or users that could be deemed \emph{inactive}. In line with other kinds of social media, bots are also a plague in social virtual worlds, with automated avatars having a noticeable impact on these~\cite{varvello2010second}.
In addition, a user might be authentic (i.e, not a bot) but be away from keyboard while their avatar is still present in the virtual world.
Determining an avatar who represents an ``authentic active user'', however, depends on an arbitrary set of characteristics and mechanisms.
Our approach, on the other hand, is relatively simplistic, but at the same time it better answers RQ2, as in any case these avatars ---active or inactive, human or bot--- occupy space and resources on Decentraland's virtual world.

Finally, our focus on quantitative methods and the lack of qualitative methods, such as user interviews and questionnaires, could be a limitation for investigating the motivations behind users' actions. We believe that using a quantitative approach based on public data to identify global patterns of actual user behavior is appropriate to answer our study's RQs. Nevertheless, future studies could also adopt a more comprehensive and multifaceted approach using mixed methods. For instance, by corroborating data analyses such as ours with surveys to obtain data about the users' self-perceived behavior, or even with ethnographic analyses~\cite{boellstorff2006ludicrous, boellstorff2020rethinking} to delve into the details of users' behaviors in social virtual worlds.

\section{Conclusions}
\label{sec:conclusions}

The concept of the Metaverse has in recent years sparked the collective imagination on novel and creative ways in which we could interact within immersive social virtual worlds and represent ourselves via our avatars.
Decentraland ---whose economy, digital assets, and governance are all based on blockchain technology--- offers a glimpse on how a potential Metaverse based on tokenomics might work in the future.
The platform has been a pioneer in developing a decentralized social virtual world in which asset ownership is radically different from previous virtual worlds. Moreover, some of its economic and artistic initiatives with worldwide renowned personalities and companies, such as the Metaverse Fashion Week and the Metaverse Music Festival, have reached the mainstream media and gathered much public attention.
At present, however, based on our results we can affirm that ICE Poker by Decentral Games (with its enticing money-earning promise) is the main driver in the dynamics of wearables and visits in Decentraland, having thus great influence on the platform in general.
This predominance of a single third-party online gambling initiative ---which has transformed itself several times due to technical, economic, and legal uncertainties--- raises doubts on if and how Decentraland's current bet on tokenomics to realize its vision of the Metaverse will ultimately pay off.

\section*{Acknowledgement}
Not applicable.

\section*{Funding}
This research received no specific grant from any funding agency in the public or private sectors.

\section*{Authors' Contributions}
All authors have participated in drafting the manuscript. All authors read and approved the final version of the manuscript.

\section*{Conflict of Interest}
The authors declare no conflict of interest.

\section*{Data Availability}
The data supporting the findings of this study are available on FigShare at: \url{https://doi.org/10.6084/m9.figshare.28596923}.

\section*{Ethical Statement}
Ethical approval was not required for the present study due to its use of anonymous and public data. In particular, the vast majority of our data was obtained from blockchain platforms with permissive software and data licenses. For these data, there is not personally identifiable information of the analyzed users, as these are anonymously identified via a unique string of hexadecimal characters that was cryptographically generated.

\section*{Declaration Of AI Usage}
No generative AI tools were used for content creation in this manuscript (e.g., drafting, rewriting, or generating ideas).

\end{multicols}

\begin{thebibliography}{99}

\bibitem{kerdvibulvech2022exploring}
Kerdvibulvech, C. (2022). Exploring the impacts of covid-19 on digital and metaverse games. In
C. Stephanidis, M. Antona, \& S. Ntoa (Eds.), \emph{HCI international 2022 posters} (pp. 561–565). Springer
International Publishing.
{}
\bibitem{bleize2019factors}
Bleize, D. N., \& Antheunis, M. L. (2019). Factors influencing purchase intent in virtual worlds: A review of
the literature. \emph{Journal of Marketing Communications}, \emph{25}(4), 403–420.
{}
\bibitem{zhou2018ownership}
Zhou, M., Leenders, M. A., \& Cong, L. M. (2018). Ownership in the virtual world and the implications for
long-term user innovation success. \emph{Technovation}, \emph{78}, 56–65.

\bibitem{cannavo2020blockchain}
Cannavo, A., \& Lamberti, F. (2020). How blockchain, virtual reality, and augmented reality are converging,
and why. \emph{IEEE Consumer Electronics Magazine}, \emph{10}(5), 6–13.
{}
\bibitem{trujillo2023toward}
Trujillo, A., \& Bacciu, C. (2023). Toward blockchain-based fashion wearables in the metaverse: The case of
decentraland. \emph{IEEE International Conference on Metaverse Computing, Networking and Applications
(MetaCom)}, 653–657.
{}
\bibitem{grimmelmann2009virtual}
Grimmelmann, J. (2009). Virtual world feudalism. \emph{Yale Law Journal Pocket Part}, \emph{118}, 126.
{}
\bibitem{axie}
Nguyen, T. T., Larsen, A. L., Doan, T., Ho, A., \& Zirlin, J. (2021). Axie infinity [Available at: \url
{https://whitepaper.axieinfinity.com/} (Accessed: 2024-7-12)]. \emph{Official Axie Infinity Whitepaper}.
{}
\bibitem{boellstorff2016whom}
Boellstorff, T. (2016). For whom the ontology turns: Theorizing the digital real. \emph{Current
Anthropology}, \emph{57}(4), 387–407.
{}
\bibitem{wood2014ethereum}
Wood, G., et al. (2014). Ethereum: A secure decentralised generalised transaction ledger. \emph{Ethereum
project yellow paper}, \emph{151}(2014), 1–32.
{}
\bibitem{hassan2021decentralized}
Hassan, S., \& De Filippi, P. (2021). Decentralized autonomous organization. \emph{Internet Policy Review},
\emph{10}(2), 1–10.
{}
\bibitem{lo2020assets}
Lo, Y. C., \& Medda, F. (2020). Assets on the blockchain: An empirical study of tokenomics.
\emph{Information Economics and Policy}, \emph{53}, 100881.
{}
\bibitem{goldberg2023metaverse}
Goldberg, M., \& Schär, F. (2023). Metaverse governance: An empirical analysis of voting within decentralized
autonomous organizations. \emph{Journal of Business Research}, \emph{160}, 113764.
{}
\bibitem{parkin2022trouble}
Parkin, S. (2022). The trouble with roblox, the video game empire built on child labour. \emph{The Guardian,
January}, \emph{9}, 2022.
{}
\bibitem{kou2023harmful}
Kou, Y., \& Gui, X. (2023). Harmful design in the metaverse and how to mitigate it: A case study of
user-generated virtual worlds on roblox. \emph{Proceedings of the 2023 ACM Designing Interactive Systems
Conference}, 175–188.
{}
\bibitem{glas2010games}
Glas, M. A. J. (2010). \emph{Games of stake: Control, agency and ownership in world of warcraft}.
Universiteit van Amsterdam [Host].
{}
\bibitem{ordano2017blockchain}
Ordano, E., Meilich, A., Jardi, Y., \& Araoz, M. (2017). A blockchain-based virtual world.
\emph{Decentraland, White Paper}.
{}
\bibitem{dowling2022fertile}
Dowling, M. (2022). Fertile land: Pricing non-fungible tokens. \emph{Finance Research Letters}, \emph{44},
102096.
{}
\bibitem{goldberg2021land}
Goldberg, M., Kugler, P., \& Schär, F. (2021). Land valuation in the metaverse: Location matters.
\emph{Available at SSRN 3932189}.
{}
\bibitem{guidi2022socialgames}
Guidi, B., \& Michienzi, A. (2022). Social games and blockchain: Exploring the metaverse of decentraland.
\emph{IEEE 42nd International Conference on Distributed Computing Systems Workshops (ICDCSW)},
199–204.
{}
\bibitem{nadini2021mapping}
Nadini, M., Alessandretti, L., Di Giacinto, F., Martino, M., Aiello, L. M., \& Baronchelli, A. (2021). Mapping
the nft revolution: Market trends, trade networks, and visual features. \emph{Scientific reports},
\emph{11}(1).
{}
\bibitem{gonzalez2022digital}
Gonzalez, P. (2022, June). \emph{Digital fashion in the metaverse} [Master’s thesis, Politecnico di Milano].
{}
\bibitem{bardzell2010virtual}
Bardzell, J., Pace, T., \& Terrell, J. (2010). Virtual fashion and avatar design: A survey of consumers and
designers. \emph{Proceedings of the 6th Nordic Conference on Human-Computer Interaction: Extending
Boundaries}, 599–602.
{}
\bibitem{boellstorff2015}
Boellstorff, T. (2015). \emph{Coming of age in second life}. Princeton University Press.
{}
\bibitem{boellstorff2020rethinking}
Boellstorff, T. (2020). Rethinking digital anthropology. In \emph{Digital anthropology} (pp. 39–60).
Routledge.
{}
\bibitem{rodriguez2022dressing}
Rodriguez Sanchez, M., \& Garcia-Badell, G. (2022). Dressing the metaverse. the digital strategies of fashion
brands in the virtual universe. \emph{Advances in Fashion and Design Research: Proceedings of the 5th
International Fashion and Design Congress, CIMODE 2022, July 4-7, 2022, Guimarães, Portugal}, 387–397.
{}
\bibitem{joy2022digital}
Joy, A., Zhu, Y., Peña, C., \& Brouard, M. (2022). Digital future of luxury brands: Metaverse, digital fashion,
and non-fungible tokens. \emph{Strategic change}, \emph{31}(3), 337–343.
{}
\bibitem{goldberg2023retailing}
Goldberg, M., Schär, F., \& Thürkauf, D. (2023). Retailing and customer engagement in the metaverse: An
empirical analysis. \emph{Available at SSRN 4547100}.
{}
\bibitem{fiedler2009quantifying}
Fiedler, I. C., \& Rock, J.-P. (2009). Quantifying skill in games—theory and empirical evidence for poker.
\emph{Gaming Law Review and Economics}, \emph{13}(1), 50–57.
{}
\bibitem{griffiths2008internet}
Griffiths, M., \& Barnes, A. (2008). Internet gambling: An online empirical study among student gamblers.
\emph{International Journal of Mental Health and Addiction}, \emph{6}, 194–204.
{}
\bibitem{fiedler2011market}
Fiedler, I., \& Wilcke, A.-C. (2011). The market for online poker. \emph{UNLV Gaming Research \& Review
Journal Available at SSRN 1747646}, \emph{16}(1).
{}
\bibitem{zaman2014motivation}
Zaman, B., Geurden, K., De Cock, R., De Schutter, B., \& Abeele, V. V. (2014). Motivation profiles of online
poker players and the role of interface preferences: A laddering study among amateur and (semi-) professionals.
\emph{Computers in Human Behavior}, \emph{39}, 154–164.
{}
\bibitem{newall2023elite}
Newall, P. W., \& Talberg, N. (2023). Elite professional online poker players: Factors underlying success in a
gambling game usually associated with financial loss and harm. \emph{Addiction Research \& Theory},
\emph{31}(6), 383–394.
{}
\bibitem{armstrong2018exploration}
Armstrong, T., Rockloff, M., Browne, M., \& Li, E. (2018). An exploration of how simulated gambling games
may promote gambling with money. \emph{Journal of Gambling Studies}, \emph{34}, 1165–1184.
{}
\bibitem{sirola2021role}
Sirola, A., Savela, N., Savolainen, I., Kaakinen, M., \& Oksanen, A. (2021). The role of virtual communities in
gambling and gaming behaviors: A systematic review. \emph{Journal of Gambling Studies}, \emph{37}(1),
165–187.
{}
\bibitem{savolainen2022online}
Savolainen, I., Sirola, A., Vuorinen, I., Mantere, E., \& Oksanen, A. (2022). Online communities and gambling
behaviors—a systematic review. \emph{Current Addiction Reports}, \emph{9}(4), 400–409.
{}
\bibitem{oleary2013online}
O’Leary, K., \& Carroll, C. (2013). The online poker sub-culture: Dialogues, interactions and networks.
\emph{Journal of Gambling Studies}, \emph{29}, 613–630.
{}
\bibitem{fang2022cryptocurrency}
Fang, F., Ventre, C., Basios, M., Kanthan, L., Martinez-Rego, D., Wu, F., \& Li, L. (2022). Cryptocurrency
trading: A comprehensive survey. \emph{Financial Innovation}, \emph{8}(1), 13.
{}
\bibitem{oecd2022lessons}
OECD. (2022). Lessons from the crypto winter. \emph{OECD Business and Finance Policy Papers}, (18).
\href {https://doi.org/https://doi.org/10.1787/199edf4f-en} {\nolinkurl
{https://doi.org/https://doi.org/10.1787/199edf4f-en}}
{}
\bibitem{houser2022navigating}
Houser, K. A., \& Holden, J. T. (2022). Navigating the non-fungible token. \emph{Utah L. Rev.}, 891.
{}
\bibitem{sharma2022s}
Sharma, T., Zhou, Z., Huang, Y., \& Wang, Y. (2022). ``It’s a blessing and a curse'': Unpacking creators’
practices with non-fungible tokens (NFTs) and their communities. \emph{arXiv preprint arXiv:2201.13233}.
{}
\bibitem{johnson2023cryptocurrency}
Johnson, B., Co, S., Sun, T., Lim, C. C., Stjepanović, D., Leung, J., Saunders, J. B., \& Chan, G. C. (2023).
Cryptocurrency trading and its associations with gambling and mental health: A scoping review.
\emph{Addictive Behaviors}, \emph{136}, 107504.
{}
\bibitem{delfabbro2021cryptocurrency}
Delfabbro, P., King, D., Williams, J., \& Georgiou, N. (2021). Cryptocurrency trading, gambling and problem
gambling. \emph{Addictive Behaviors}, \emph{122}, 107021.
{}
\bibitem{andrade2023safer}
Andrade, M., Sharman, S., Xiao, L. Y., \& Newall, P. W. (2023). Safer gambling and consumer protection
failings among 40 frequently visited cryptocurrency-based online gambling operators. \emph{Psychology of
Addictive Behaviors}, \emph{37}(3), 545.
{}
\bibitem{de2022play}
De Jesus, S. B., Austria, D., Marcelo, D. R., Ocampo, C., Tibudan, A. J., \& Tus, J. (2022). Play-to-earn: A
qualitative analysis of the experiences and challenges faced by axie infinity online gamers amidst the covid-19
pandemic. \emph{International Journal of Psychology and Counseling}, \emph{1}(12), 291–424.
{}
\bibitem{delic2024profiling}
Delic, A. J., \& Delfabbro, P. H. (2024). Profiling the potential risks and benefits of emerging ``play to earn''
games: A qualitative analysis of players’ experiences with axie infinity. \emph{International Journal of Mental
Health and Addiction}, \emph{22}(1), 634–647.
{}
\bibitem{hong2023wean}
Hong, W., Jing, Y., Liu, R.-D., Ding, Y., \& Wang, J. (2023). Wean your child off video games: Using external
rewards to undermine intrinsic motivation to play interesting video games. \emph{Current Psychology},
\emph{42}(21), 17746–17756.
{}
\bibitem{miikkulainen2023digital}
Miikkulainen-Gilbert, H. (2023). \emph{Digital transformation: How can it provide the most value to fashion
brands while enabling sustainable change?} [Master’s thesis, Arcada University of Applied Sciences].
{}
\bibitem{dolata2023metaverse}
Dolata, M., \& Schwabe, G. (2023). What is the metaverse and who seeks to define it? mapping the site of
social construction. \emph{Journal of Information Technology}, \emph{38}(3), 239–266.
{}
\bibitem{echauri2025tech}
Echauri, G. (2025). Tech trends: Wired’s treatment of emerging technologies (2021-2023). \emph{Journalism},
\emph{26}(6), 1270–1287.
{}
\bibitem{zborowska2012regulation}
Zborowska, N., Kingma, S. F., \& Brear, P. (2012). Regulation and reputation: The gibraltar approach. In
\emph{Routledge international handbook of internet gambling} (pp. 84–99). Routledge.
{}
\bibitem{trautman2022ftx}
Trautman, L. J., Foster, I., \& Larry, D. (2022). The ftx crypto debacle: Largest fraud since madoff?
\emph{University of Memphis Law Review, Forthcoming}.
{}
\bibitem{kerr2018can}
Kerr, J. (2018). How can legislators protect sport from the integrity threat posed by cryptocurrencies?
\emph{The International Sports Law Journal}, \emph{18}(1), 79–97.
{}
\bibitem{fischer2021poker}
Fischer, J. M. (2021). Is poker legal or illegal in california? \emph{UNLV Gaming LJ}, \emph{12}, 27.
{}
\bibitem{macegoniuk2021british}
Macegoniuk, A. (2021). \emph{How british island tax havens accommodate financial criminals? historical
perspective 1956–2020} [Master’s thesis, Univeristy of Oulu].
{}
\bibitem{zaucha2024racial}
Zaucha, T. (2024). Racial capitalism, platform capitalism, and nfts. In \emph{Non-fungible tokens}
(pp. 139–154). Routledge.
{}
\bibitem{xiao2024loot}
Xiao, L. Y. (2024). Loot box state of play 2023: Law, regulation, policy, and enforcement around the world.
\emph{Gaming Law Review}, \emph{28}(10), 450–483.
{}
\bibitem{varvello2010second}
Varvello, M., \& Voelker, G. M. (2010). Second life: A social network of humans and bots. \emph{Proceedings
of the 20th international workshop on network and operating systems support for digital audio and video},
9–14.
{}
\bibitem{boellstorff2006ludicrous}
Boellstorff, T. (2006). A ludicrous discipline? ethnography and game studies. \emph{Games and culture},
\emph{1}(1), 29–35.
\end{thebibliography}
\end{document}